

\hsize=35truepc\hoffset=2pc
\vsize=54truepc

\font\titlefont=cmcsc10 at 14pt
\font\namefont=cmr12
\font\placefont=cmsl12
\font\abstractfont=cmbx12
\font\secfont=cmr12

\font\twrm=cmr12         \font\twmi=cmmi12
  \font\twsy=cmsy10 at 12pt  \font\twex=cmex10 at 12pt
  \font\twit=cmti12          \font\twsl=cmsl12
  \font\twbf=cmbx12
\font\nnrm=cmr9          \font\nnmi=cmmi9
  \font\nnsy=cmsy9           \font\nnex=cmex10 at 9pt
  \font\nnit=cmti9           \font\nnsl=cmsl9
  \font\nnbf=cmbx9

\font\sxrm=cmr6          \font\sxmi=cmmi6             
  \font\sxsy=cmsy6

\def\nnpt{\def\rm{\fam0\nnrm}%
  \textfont0=\nnrm \scriptfont0=\sxrm \scriptscriptfont0=\fiverm
  \textfont1=\nnmi \scriptfont1=\sxmi \scriptscriptfont1=\fivei
  \textfont2=\nnsy \scriptfont2=\sxsy \scriptscriptfont2=\fivesy
  \textfont3=\nnex \scriptfont3=\nnex \scriptscriptfont3=\nnex
  \textfont\itfam=\nnit \def\it{\fam\itfam\nnit}%
  \textfont\slfam=\nnsl \def\sl{\fam\slfam\nnsl}%
  \textfont\bffam=\nnbf \def\bf{\fam\bffam\nnbf}%
  \normalbaselineskip=11pt
  \setbox\strutbox=\hbox{\vrule height8pt depth3pt width0pt}%
  \let\big=\nnbig \let\Big=\nnBig \let\bigg=\nnbigg \let\Bigg=\nnBigg
  \normalbaselines\rm}

\def\tnpt{\def\rm{\fam0\tenrm}%
  \textfont0=\tenrm \scriptfont0=\sevenrm \scriptscriptfont0=\fiverm
  \textfont1=\teni  \scriptfont1=\seveni  \scriptscriptfont1=\fivei
  \textfont2=\tensy \scriptfont2=\sevensy \scriptscriptfont2=\fivesy
  \textfont3=\tenex \scriptfont3=\tenex    \scriptscriptfont3=\tenex
  \textfont\itfam=\tenit \def\it{\fam\itfam\tenit}%
  \textfont\slfam=\tensl \def\sl{\fam\slfam\tensl}%
  \textfont\bffam=\tenbf \def\bf{\fam\bffam\tenbf}%
  \normalbaselineskip=12pt
  \setbox\strutbox=\hbox{\vrule height8.5pt depth3.5pt width0pt}%
  \let\big=\tenbig \let\Big=\tenBig \let\bigg=\tenbigg \let\Bigg=\tenBigg
  \normalbaselines\rm}

\def\twpt{\def\rm{\fam0\twrm}%
  \textfont0=\twrm \scriptfont0=\nnrm \scriptscriptfont0=\sevenrm
  \textfont1=\twmi \scriptfont1=\nnmi \scriptscriptfont1=\seveni
  \textfont2=\twsy \scriptfont2=\nnsy \scriptscriptfont2=\sevensy
  \textfont3=\twex \scriptfont3=\twex \scriptscriptfont3=\twex
  \textfont\itfam=\twit \def\it{\fam\itfam\twit}%
  \textfont\slfam=\twsl \def\sl{\fam\slfam\twsl}%
  \textfont\bffam=\twbf \def\bf{\fam\bffam\twbf}%
  \normalbaselineskip=14pt
  \setbox\strutbox=\hbox{\vrule height10pt depth4pt width0pt}%
  \let\big=\twbig \let\Big=\twBig \let\bigg=\twbigg \let\Bigg=\twBigg
  \normalbaselines\rm}

\catcode`\@=11
\def\nnbig#1{{\hbox{$\left#1\vbox to7.5pt{}\right.\n@space$}}}
\def\nnBig#1{{\hbox{$\left#1\vbox to10.5pt{}\right.\n@space$}}}
\def\nnbigg#1{{\hbox{$\left#1\vbox to13.5pt{}\right.\n@space$}}}
\def\nnBigg#1{{\hbox{$\left#1\vbox to16.5pt{}\right.\n@space$}}}

\def\tenbig#1{{\hbox{$\left#1\vbox to8.5pt{}\right.\n@space$}}}
\def\tenBig#1{{\hbox{$\left#1\vbox to11.5pt{}\right.\n@space$}}}
\def\tenbigg#1{{\hbox{$\left#1\vbox to14.5pt{}\right.\n@space$}}}
\def\tenBigg#1{{\hbox{$\left#1\vbox to17.5pt{}\right.\n@space$}}}

\def\twbig#1{{\hbox{$\left#1\vbox to10.5pt{}\right.\n@space$}}}
\def\twBig#1{{\hbox{$\left#1\vbox to14pt{}\right.\n@space$}}}
\def\twbigg#1{{\hbox{$\left#1\vbox to17.5pt{}\right.\n@space$}}}
\def\twBigg#1{{\hbox{$\left#1\vbox to21pt{}\right.\n@space$}}}
\catcode`\@=12

\def\center{\parindent=0pt\leftskip=1in plus 1fill\rightskip=1in plus 1fill}

\def\title#1\par{{\center\baselineskip=16pt
  \namefont{#1}\par}}

\long\def\author #1// #2// #3// #4//{{\parskip=0pt\center\namefont{#1}\medskip
  \placefont#2\par#3\par#4\par}}

\long\def\abstract#1//{%
  \centerline{\abstractfont Abstract}\medskip
  {\baselineskip=12pt\advance\leftskip by 3pc\advance\rightskip by
3pc\parindent=10pt
  \def\enspace{\kern.3em}
  \noindent #1\par}}

\def\body{\twpt}

\def\sectitle{\center\baselineskip=14pt\lineskiplimit=1pt\secfont}

\def\beginsec#1\par{%
  \ifdim0.8\vsize<\pagetotal\ifdim\pagetotal<\pagegoal\vfil\eject\fi\else
    \removelastskip\medskip\vskip\parskip\fi
  {\sectitle #1\hfilneg\ \par}%
  \nobreak\noindent}

\def\references{\leftskip=\parindent\parindent=0pt
  \vskip0pt plus.07\vsize\penalty-250\vskip0pt plus-.07\vsize
  \removelastskip\bigskip\bigskip\vskip\parskip
  \centerline{\secfont References}\bigskip}

\newcount\refno\refno=0
\long\def\rfrnc#1//{\advance\refno by 1
  $ $\llap{\hbox to\leftskip{[\the\refno]\enspace\hfil}}#1\par\medskip}

\skip\footins=0.2in
\dimen\footins=4in
\catcode`\@=11
\def\vfootnote#1{\insert\footins\bgroup\nnpt
  \interlinepenalty=\interfootnotelinepenalty
  \splittopskip=\ht\strutbox
  \splitmaxdepth=\dp\strutbox \floatingpenalty=20000
  \leftskip=0pt \rightskip=0pt \spaceskip=0pt \xspaceskip=0pt
  \setbox1=\hbox{*}\parindent=\wd1\let\enspace=\null
  \hangafter1\hangindent\parindent\textindent{#1}\footstrut
  \futurelet\next\fo@t}
\catcode`\@=12

\def\footnoterule{\kern-3pt \hrule width2truein \kern 3.6pt}

\pretolerance=300\tolerance=400\hyphenpenalty=100

\def\Frac#1#2{{\raise.2ex\hbox{$\scriptstyle#1$}%
  \kern-.1em\scriptstyle/
  \kern-.1em\lower.2ex\hbox{$\scriptstyle#2$}}}
\def\frac#1#2{{\raise.2ex\hbox{$\scriptscriptstyle#1$}%
  \kern-.1em\scriptscriptstyle/
  \kern-.1em\lower.2ex\hbox{$\scriptscriptstyle#2$}}}


\newdimen\jbarht\jbarht=.2pt
\newdimen\vgap\vgap=1pt
\newcount\shiftfactor\shiftfactor=12

\catcode`\@=11
\def\jbarout{\setbox1=\vbox{\offinterlineskip
  \dimen@=\ht0 \multiply\dimen@\shiftfactor \divide\dimen@ 100
    \hsize\wd0 \advance\hsize\dimen@
  \hbox to\hsize{\hfil
    {\multiply\dimen@-2 \advance\dimen@\wd0
    \vrule height\jbarht width\dimen@ depth0pt}%
    \hskip\dimen@}%
  \vskip\vgap\box0\par}\box1}

\catcode`\@=12


\def\cut#1{\setbox0=\hbox{$#1$}\setbox1=\hbox to \wd0{\hss{\it/\/}\hss}
  \box1\hskip-\wd0\box0}                   


\def\simlt{\mathop{\lower.4ex\hbox{$\buildrel<\over\sim$}}}
\def\simgt{\mathop{\lower.4ex\hbox{$\buildrel>\over\sim$}}}

\font\secfont=cmbx12
\font\titlefont=cmbx12 at 14pt
\vsize=52truepc\voffset=1pc
\hsize=37truepc\hoffset=1pc
\pageno=1
\footline{\ifnum\pageno>1\hss\lower.25in\hbox{\tenrm\folio}\hss\fi}

\widowpenalty=1000
\advance\baselineskip by 2pt

\bgroup
\rightline{\it Submitted to Physics Review Letters}
\rightline{MIU-THP-93/63}
\rightline{March, 1993}
\vskip.6in
\title{\titlefont Quantum Memory Effects in the Measurement of Observables with
a
Continuous Spectrum} \vskip.2in
\author
Kai J. Dr\"uhl//
Department of Physics//
Maharishi International University//
Fairfield, Iowa 52557//

\vskip.4in

\abstract\advance\leftskip by 20pt\advance\rightskip by 20pt
In the measurement of a continuous observable $Q$, the pure components of the
reduced
state do, in general, depend on the initial state.  For measurements which
attempt to
localize the measured system in a certain region $R$, the localized wave
functions are
proportional to the original wave function outside of $R$.  This ``quantum
memory"
effect shows that it is not possible to perfectly localize a quantum particle.
\par//
\vskip.2in \egroup

\body
\parindent=25pt
\parskip\medskipamount
\advance\baselineskip by 6pt

In this paper we address the question of how to measure a quantum-mechanical
observable
with a continuous spectrum. Ever since the pioneering work of von Neumann [1],
it has
been known that for the measurement of an observable $Q$, the interaction
energy
between the observed system $S$ and the measuring apparatus $A$ has to be a
function of
$Q$ [2,3]. Von Neumann considered an interaction term which is linear in a
position-like variable $Q$ [1]. More recently Zurek studied interactions
between $S$
and $A$ in terms of a preferred ``pointer basis" of $A$ [2]. Haake and Walls
studied
the case of an oscillator coupled to an infinite reservoir [3], following the
earlier
work of Ullersma [4]. Unruh and Zurek recently studied state reduction in a
similar
model, including the case of a free particle, and presented explicit
calculations of
reduced density matrices for states with Gaussian characteristic functions [5].

These studies show that the variance of $Q$ in the pure components of
the reduced density matrix $\rho_S$ is much less than in the initial state of
$S$, in
accord with the idea that an approximate measurement of $Q$ has been performed
by $A$.
However, the explicit calculations of $\rho_S$ also show that in general the
set of
eigenfunctions of $\rho_S$ depends quite strongly on the initial state.
For a measurement, on the other hand, we require that the set of
eigenstates of $\rho_S$ be independent of the initial state, and be
identical to the set of eigenstates of the observable to be measured. In
this sense, the linear models studied so far do not provide models for a
measurement of $Q$.

In the following, we show that measurements which determine the value of a
continuous observable $Q$ can be approximately realized by couplings of the
form
$\lambda_S(Q)\cdot V_A$, where $\lambda_S$ is a strongly localized function
which vanishes rapidly outside a finite interval $I$, and $V_A$ is an apparatus
operator.  However, in this case the localized wave functions do not vanish,
but are
proportional to the original wave function outside\break of $I$.

Following von Neumann, we assume that the Hamiltonian of system $S$ plus
apparatus $A$ takes the form
$$H=\lambda_S(Q)\cdot V_A+H_A\eqno{(1)}$$
and that the initial state $\rho (0)$ of system $S$ plus apparatus $A$ at time
$t=0$ is the product of a pure state $\left\vert
\psi_S\rangle\langle\psi_S\right\vert$ of
$S$ with a state $\rho_A$ of the apparatus. Due to the $S$-$A$ interaction
term, at later
times $t>0$ the state $\rho (t)$ is no longer a product, and the reduced state
$\rho_S (t)={\rm trace}_A\ \rho (t)$ of $S$ is no longer a pure state. This
state
reduction models the measuring process.

The Hamiltonian (1) does not contain a part for the free motion of $S$.  This
corresponds to the assumption that the system $S$ does not change significantly
over a
time period $t$ required for the measurement.

We further assume that $[V_A, H_A]=0$, and that the initial state of $A$ is
invariant
under the free motion of $A$.  The matrix elements of $\rho_S(t)$ in the basis
$\{\vert q\rangle\}$ of eigenstates of $Q$ are then (we choose units
such that $\hbar=1$)
$$\eqalignno{
\langle q'\left\vert\rho_S(t)\right\vert q''\rangle&={\rm trace}_A\ \langle
q'\vert e^{-itH} (\left\vert\psi_S\rangle\langle \psi_S\right\vert
\otimes\rho_A) e^{itH}\vert q''\rangle\cr
&=\langle q'\vert \psi_S\rangle\langle \psi_S\vert q''\rangle\ {\rm trace}_A\
\bigl(e^{-it\lambda_S (q')V_A}\rho_A e^{it\lambda_S(q'')V_A}\bigr)\cr
&=\psi_S(q')\psi_S(q'')^*f_A\Bigl(t\bigl(\lambda_S(q')-\lambda_S(q'')\bigr)\Bigr).&(2)}$$
\vfill
\eject
The overlap function $f_A(t)={\rm trace}_A\ \bigl(\rho_Ae^{-it V_A}\bigr)$ has
the property $f_A (0)=1,$ and we assume that $\lim_{t\to\infty}f_A(t)=0$.
We now apply equation (2) to three different cases of interest.

Firstly, assume that $\lambda_S(Q)\equiv Q$, and that $Q$ has a discrete
spectrum
$\{q_\alpha\}$, such that
$\vert q_\alpha-q_\beta\vert\ge\Delta q>0\ {\rm for}\ \alpha\not=\beta$.
Then with $\theta_{\alpha\beta}=t(q_\alpha - q_\beta)$ and
$\vert\theta_{\alpha\beta}\vert<t\Delta q$ we have for the matrix elements of
$\rho_S$ in the basis $\{\vert q_\alpha\rangle\}$:
 $$\eqalignno {
\langle q_\alpha\vert\rho_S(t)\vert q_\alpha\rangle &= \vert\langle
q_\alpha\vert \psi_S\rangle\vert^2\cr
\langle q_\alpha\vert\rho_S(t)\vert q_\beta\rangle &= \langle q_\alpha\vert
\psi_S\rangle\langle\psi_S\vert q_\beta\rangle f(\theta_{\alpha\beta})\cr
&\to 0 \ \ {\rm for}\ t\to\infty,\ \alpha\not=\beta.&(3)}$$
The off-diagonal elements go to zero uniformly in $\alpha$ and $\beta$ for
$t\to\infty$.  This shows that $\rho_S(t)$ converges to its diagonal part in
the limit
$t\to\infty$.

Secondly, we consider $\lambda_S(Q)\equiv Q$ with a continuous observable $Q$,
and a
Gaussian overlap function.
$$f_A\bigl(t(q'-q'')\bigr)=\exp\Bigl(-{1\over
4}\kappa^2_A(t)\bigl(q'-q''\bigr)^2\Bigr)$$
If the initial state $\vert\psi_S\rangle$ is the
vacuum state of the shifted number operator $N={1\over
2}\Bigl(\kappa^{-2}_SP^2+\kappa^2_S \bigl({Q-q_0}\bigr)^2\Bigr)-{1\over 2}$,
then
$$\eqalignno{
\psi_S(q)=\langle q\vert \psi_S\rangle &= {1\over \sqrt\pi}\exp\biggl(-{1\over
2}\kappa^2_S ({q-q_0)}^2\biggr),\ \ {\rm and}\cr
\langle q'\vert\rho_S\vert q''\rangle &={1\over\pi}\exp\biggl(-{1\over
2}\kappa^2_S\biggl(({q'-q_0})^2+({q''-q_0})^2-{1\over
4}\kappa^2_A(q'-q'')^2\biggr)&(4)}$$

The state $\rho_S$ is a canonical ensemble
$\rho_S=\bigl(1-e^{-\beta}\bigr) e^{-\beta N{_R}}$
whose pure components are the eigenstates of the squeezed, shifted number
operator $N_R$
$$ N_R={1\over 2}\Bigl(\kappa^{-2}_R P^2+\kappa^2_R ({Q-q_0})^2\Bigr)-{1\over
2},\ \  {\rm with}$$
$$\kappa^2_R=\kappa_S\sqrt {\kappa^2_S+\kappa^2_A}\ \ {\rm and}\ \
e^{-\beta}={\Bigl(\kappa^2_R-\kappa^2_S\Bigr)\over\Bigl(\kappa^2_R+\kappa^2_S\Bigr)}.\eqno{(5)}$$
For significant state reduction, $\kappa^2_A\gg\kappa^2_S$, and
$\kappa_R^2\approx\kappa_S\kappa_A$ still depends strongly on the initial state
through
$\kappa_S$ and $q_0$.  The eigenvalues of $\rho_S$ are independent of $q_0$,
the location
of the maximum of $\psi_S$.

Finally, we consider the case of continuous $Q$ with strongly localized
coupling
function $\lambda_S(q)$.  In the physically unrealistic case where
$\lambda_S=1$
inside and $\lambda_S=0$ outside of a finite interval $I$, the eigenstates of
$\rho_S$
are wave functions with support inside or outside of $I$.  In this situation,
the
observable being measured is the spectral projection operator associated with
eigenvalues of $Q$ in $I$.

For physically realizable interactions, the function $\lambda_S(q)$ will not be
a step
function, but will go to zero for $\vert q\vert\to\infty$.  An example is the
coupling
between two levels of an atom through the electric field of an external
electron,
which depends on the distance between electron and atom.

In such a situation, if the wave function $\psi_S(q)$ does not
vary strongly in the region, where $\lambda_S(q)$ is appreciably different
from zero, we expect the pure components of the reduced state to be independent
of the
initial state vector $\vert\psi_S\rangle$. More specifically, we assume that at
sufficiently large time $t$, there exists a value $q_0$, such that
$$f_A\bigl(t\lambda_S(q)\bigr)\approx 1\ {\rm for}\ \vert q\vert>q_0,\
f_A\bigl(t\lambda_S(0)\bigr)\approx 0\ \ {\rm and}\ \  \psi_S(q)\approx
\psi_S(0)\ {\rm
for}\ \vert q\vert <q_0.\eqno{(6)}$$
 From Eq.~(2) we find that the eigenfunctions of $\rho_S$ with
non-zero eigenvalues are proportional to $\psi_S(q)$ for $\vert q\vert>q_0$. In
this
subspace of wave functions, the density matrix $\rho_S$ still depends on
$\psi_S(0)$
or on the probability $p_0=2q_0\vert\psi_S(0)\vert^2$ of finding the system in
the
interval $I=[-q_0, q_0]$.  For small values of $p_0$, a perturbation expansion
in
$\psi_0(0)\equiv \psi$ gives the following results.  The eigenfunctions of
$\rho_S$ take
the form:
$$\eqalignno{
\psi_N(q)=\psi_S(q) f_A\bigl(t\lambda_S(q)\bigr) &+ 0(\psi^2),\cr
\psi_i(q)=\psi_{i0}(q)+\lambda_{iM}\psi_M(q)&+0(\psi^2),&(7)}$$
where $\psi_{i0}(q)\approx 0$ and $\psi_M(q)\approx\psi_N(q)\approx\psi_S(q)$
for
$\vert q\vert>q_0$.

The function $\psi_N$ has zero amplitude at $q=0$, and corresponds to a
negative
result for the attempt to localize the system $S$ in $I$, which this
measurement
represents.  The functions $\psi_i$, on the other hand, represent different
modes of
localization of $S$ within $I$.  The coefficients $\lambda_{iM}$ are non-zero
and
proportional to $\psi_S(0)$.  Thus, the localized wave functions do not vanish
outside
of $I$, but are proportional to the original wave function there.

To illustrate these findings, we have diagonalized $\rho_S$ numerically by
discretizing
the matrix elements in the interval $I$. In the example shown here, the
coupling function
is $\lambda_S(q)=\exp(-q^2)$, overlap function $f(\theta)=\exp(-\theta^2)$ and
time
$t=4$. The resulting eigenfunctions were scaled by the square root of their
eigenvalues, and the scaled eigenfunctions with largest eigenvalues are shown
in Figure 1
for values of $p_0=0.01$ and $p_0=0.36$. As can be seen, the eigenfunctions for
both
cases are almost identical. The non-localized wave function is very close to
$\psi_N$
of Eq.~(7) even for large values of $p_0$. In addition, there exist about nine
localized wave functions with significant eigenvalues.  For $p_0=0.36$, the
asymptotic
tail of the localized eigenfunctions for large $q$, resulting from the
coefficient
$\lambda_{iM}$, is clearly visible. This relatively large number of different
modes of
localization is due to the fact that we have treated the apparatus $A$ as
completely
unobservable. For a more realistic model, in which $A$ is partly observable, a
smaller
number of modes may result.

The results obtained here are interesting in that they show a fundamental
difference in measurability between continuous observables with linear
couplings and observables with nonlinear, localized couplings. Examples of
the first category with linear couplings only are field variables, such as
the electric and magnetic fields. Such variables cannot be measured in the
strict sense of the word, since the pure components of their reduced state
depend upon the initial state. In the second example above, the corresponding
eigenfunctions are all centered at the same location $q_0$ as the original wave
function
$\psi_S$, while the eigenvalues are independent of $q_0$.  The first example
shows that
this is due exclusively to the continuous nature of $Q$.  While in the discrete
case,
the density matrix elements converge uniformly to diagonal form, in the
continuous case
convergence is non-uniform.  For a continous variable, there does not exist a
diagonal
form of the density matrix, and example two shows that the reduced state
converges to
zero in the norm for large times.  We also note that our result (4) for the
reduced
matrix elements, although obtained under somewhat restrictive conditions on the
total
Hamiltonian, is in agreement with the results obtained by Unruh and Zurek [5].
They
find that the term involving $\kappa^2_A$ is dominant for times
$t\sim\Gamma^{-1}$, where
$\Gamma$ is the cut-off frequency of the infinite reservoir of oscillations
representing $A$ in their work.
\vfill
\eject

Finally, example three demonstrates the possibility of measuring continuous
observables
such as the position of a charged particle, which couple to a measuring
apparatus
through localized functions of position, e.g. their electric field.  Such
couplings allow
one to decide the question whether or not the particle is approximately
localized at a
given point, with corresponding eigenfunctions that are almost independent of
the initial
state. In this sense, these observables can be measured indirectly.  However,
in this
case, the asymptotic behavior of the localized eigenfunctions at large
distances reveals
a quantum-memory effect, or incomplete localization of the particle by this
type of
measurement.  Thus, if the particle had been found localized at $q=0$, an
immediately
following localization experiment of the same type, but at a different location
$q=q_1$,
would yield a positive result with a probability proportional to
$\vert\psi_S(0)\vert^2\ \vert\psi_S(q_1)\vert^2$.
\vfill
\eject

\references

\rfrnc J. von Neumann, {\it Mathematical Foundations of Quantum Mechanics,
Chap.~VI},
Princeton University Press, Princeton (1955).//

\rfrnc W.H. Zurek, {\it Phys.~Rev.}\/ {\bf D24} (1981) 1516; {\bf 26} (1982)
1862.//

\rfrnc F. Haake and D.F. Walls, {\it Phys.~Rev.}\/ {\bf A36} (1987) 730.//

\rfrnc P. Ullersma, {\it Physica (Utrecht)}\/ {\bf 32} (1966) 27; 56; 74; 90.//

\rfrnc W.G. Unruh and W.H. Zurek, {\it Phys.~Rev.}\/ {\bf D40} (1989) 1071.//
\vfill
\eject
\centerline {\bf Figure Caption}
\vskip .2in
\parindent=0pt
\hangindent=60pt\hangafter=1
Figure 1: The most important localized eigenfunctions and the non-localized
eigenfunction of the reduced density matrix are shown for two different values
of the
local probability $p_0$.  Solid curves correspond to $p_0=0.01$, and dashed
curves to
$p_0=0.36$.  The functions have been scaled by the square root of their
eigenvalue.

\end